\documentclass[letterpaper,11pt]{article}

\usepackage{amssymb}

\def\beq{\begin{equation}}
\def\eeq{\end{equation}}
\def\bea{\begin{eqnarray}}
\def\eea{\end{eqnarray}}

\begin{document}

\pagestyle{empty}

\hfill {\tt hep-th/0405056}

\begin{center}

\vspace*{2cm}

\noindent
{\Large \bf Self-dual strings in six dimensions: \\
Anomalies, the $ADE$-classification, \\ and the world-sheet $WZW$-model}
\vskip 2truecm

{\Large M{\aa}ns Henningson} \\ 
\vskip 1truecm
{\it Department of Theoretical  Physics\\ Chalmers University of
  Technology and G\"oteborg University\\ SE-412 96 G\"{o}teborg,
  Sweden \\ {\tt mans@fy.chalmers.se}}\\
\end{center}

\vskip 3cm
\noindent{\bf Abstract:}

\noindent
We consider the $(2, 0)$ supersymmetric theory of tensor multiplets
and self-dual strings in six space-time dimensions.  Space-time
diffeomorphisms that leave the string world-sheet invariant appear as
gauge transformations on the normal bundle of the world-sheet.  The
naive invariance of the model under such transformations is however explicitly
broken by anomalies: The electromagnetic coupling of the string to the
two-form gauge field of the tensor multiplet suffers from a classical
anomaly, and there is also a one-loop quantum anomaly from the chiral
fermions on the string world-sheet.  Both of these contributions are
proportional to the Euler class of the normal bundle of the string 
world-sheet, and consistency
of the model requires that they cancel.  This imposes strong
constraints on possible models, which are found to obey an
$ADE$-classification.  We then consider the decoupled world-sheet
theory that describes low-energy fluctuations (compared to the scale
set by the string tension) around a configuration with a static,
straight string. The anomaly structure determines this to be a
supersymmetric version of the level one
Wess-Zumino-Witten model based on the group $({\mathbb R} \times SU (2))
/ {\mathbb Z}_{2}$.

\newpage

\pagestyle{plain}

\section{Introduction} 
In a recent paper \cite{Arvidsson-Flink-Henningson}, we constructed a $(2, 0)$ supersymmetric theory of tensor multiplets and self-dual strings in six dimensions. The model is formulated in terms of fields defined over a six-dimensional Minkowski space-time $M$, and fields defined over a three-dimensional Dirac-membrane world-volume $D$, the boundary $\partial D$ of which equals the string world-sheet $\Sigma$. The fields over $M$ are scalars $\phi$, chiral spinors $\psi$, and a two-form gauge-field $b$ with gauge-invariant field-strength $h = d b$. The fields over $D$ are a Minkowski space vector $X$ and a Minkowski space anti-chiral spinor $\Theta$. The $(2, 0)$ supersymmetry algebra contains an $SO (5)$ $R$-symmetry, under which $\phi$ transforms in the vector representation, $\psi$ and $\Theta$ transform in the spinor representation, and $b$ and $X$ are invariant. All fields obey certain reality conditions. 

This model is invariant under a local `$\kappa$-symmetry', by means of which most of the fields $X$ and $\Theta$ may be gauged away. Although this was not explicitly shown in the paper, the remaining theory could then be described in terms of the fields $\phi$, $\psi$, and $b$ over $M$ together with certain fields $X^\perp$, $\Theta^+$, and $\Theta^-$ defined over the string world-sheet $\Sigma$. The latter can be understood as follows: The world-sheet field $X$ defines an embedding $X : \Sigma \rightarrow M$. Consider the pullback ${}^*(TM) = X^* (TM)$ of the tangent bundle $TM$ of $M$ to the string world-sheet $\Sigma$ by this map. (In this paper, a raised star ${}^*$ will always denote pullbacks from $M$ to $\Sigma$ by $X$.) This bundle splits as ${}^* (TM) = T \Sigma \oplus N$, where $T \Sigma$ is the tangent bundle of $\Sigma$, and $N$ is its orthogonal complement, i.e. the normal bundle of $\Sigma$ in $M$. Furthermore, the pullback ${}^* (\phi)$ of the scalar field $\phi$ to $\Sigma$ defines an embedding of $\Sigma$ in a five-dimensional real vector space, the normal bundle of which we denote as $E$. Associated to the $SO (1, 1)$ bundle $T \Sigma$, there are chiral spinor bundles $\Sigma^+$ and $\Sigma^-$, where the superscript denotes the chirality. Similarly, associated to the $SO (4)$ bundles $N$ and $E$, there are chiral spinor bundles $N^+$, $N^-$ and $E^+$, $E^-$ respectively. All these bundles over $\Sigma$ are endowed with natural connections induced from the embedding of $\Sigma$. The fields over the world-sheet $\Sigma$ are a bosonic section $X^\perp$ of $N$, describing transverse fluctuations of $\Sigma$, together with fermionic sections $\Theta^+$ and $\Theta^-$ of $\Sigma^+ \otimes N^+ \otimes E^+$ and $\Sigma^- \otimes N^- \otimes E^-$ respectively. 

The model can be given a Lagrangian formulation with an action
\bea
S & = & \frac{1}{4 \pi \lambda^2} \int_M \left({\rm Vol}_M \left( \partial \phi \partial \phi + \bar{\psi} \slash{\!\!\!\! \partial} \psi \right) + h^{\rm tot} \wedge * h^{\rm tot} \right) \cr
& & + \int_\Sigma \left( {\rm Vol}_\Sigma \sqrt{{}^* (\phi \phi)} \left( D X^\perp D X^\perp + \bar{\Theta}^+ \slash{\!\!\!\! D} \Theta^+ + \bar{\Theta}^- \slash{\! \! \! \! D} \Theta^- \right) + e \, {}^* (b) \right) + \ldots , \label{action}
\eea
where the omitted terms will not be important in the present paper. Here ${\rm Vol}_M$ and ${\rm Vol}_\Sigma$ are the volume form on $M$ and the induced volume form on $\Sigma$ respectively, $*$ denotes the Hodge duality operator, $e$ is the electric charge of the string, and $\lambda$ is a coupling constant. The world-sheet derivative $D$ acting on sections of various bundles is constructed using the appropriate connection. The total field strength $h^{\rm tot}$ obeys a modified Bianchi identity
\beq
d h^{\rm tot} = 2 \pi q \, \delta_\Sigma , \label{Bianchi}
\eeq
where $q$ is the magnetic charge of the string, and $\delta_\Sigma$ is the Poincar\'e dual four-form of the string world-sheet $\Sigma$. The anti self-dual part $h_- \equiv \frac{1}{2} \left( h + * h \right)$ of the field strength $h$, which is not part of the tensor multiplet, decouples from the rest of the theory provided that we choose the coupling constant so that $\lambda^2 = q / e$.

The group of diffeomorphisms of $M$ is of course spontaneously broken by the string to the subgroup that leaves the string world-sheet $\Sigma$ invariant. From a world-sheet perspective, this subgroup appears as an $SO (4)$ gauge group acting on the normal bundle $N$. Naively, the action (\ref{action}) is invariant under such transformations. However, because of the modified Bianchi identity (\ref{Bianchi}), the electric coupling term $e \, {}^* (b)$ in fact transforms anomalously already at the classical level. \footnote{The construction in \cite{Arvidsson-Flink-Henningson} involved regulating the theory by introducing a perturbation $\Sigma^\prime$ of the world-sheet $\Sigma$. This is rather analogous to the framed knot discussed in \cite{Witten89}. As pointed out to me by E.~Witten, the necessity to regulate the theory in this way is a symptom of the underlying anomaly.} Furthermore, the chiral fermions $\Theta^+$ and $\Theta^-$ transform in complex representations of $SO (4)$, so the transformation of the quantum effective action acquires further anomalous terms at one-loop level. In the next section, we will describe these contributions to the total anomaly more carefully. 

At this point, it should be stressed that we are considering a theory in flat six-dimensional Minkowski space, and we are not attempting to gauge the $SO (5)$ $R$-symmetry. It is well known that in e.g. the $(2, 0)$ supersymmetric world-volume theory on five-branes embedded in eleven-dimensional $M$-theory, the local diffeomorphism and $SO (5)$ symmetries a priori suffer from anomalies. This would of course spoil the consistency of the theory. However, as shown in \cite{Freed-Harvey-Minasian-Moore}, thanks to delicate cancellations between the six-dimensional world-volume anomalies, anomaly inflow terms from the eleven-dimensional bulk, and a subtle contribution that originates from the Chern-Simons interaction of eleven-dimensional $M$-theory, the total anomalies in fact vanish.  The six-dimensional theory is thus consistent when embedded into eleven-dimensional $M$-theory. However, it is not consistent in itself when coupled to an arbitrary curved background metric and $SO (5)$ connection. But in the situation that we are considering here, i.e. a flat space and a global $SO (5)$ $R$-symmetry, there is no problem, and we need not be concerned with gravitational and gauge anomalies. The particular string world-sheet anomalies that are the main subject of the present paper have, to the best of our knowledge, not been considered previously.  (However, some related issues were discussed already in \cite{Brax-Mourad}, and world-sheet anomalies on solitonic string solutions in a five-dimensional field theory are considered in \cite{Boyarsky-Harvey-Ruchayskiy}.)

Consistency of the model requires that the total anomaly cancels, and as we will see in section three, this imposes strong restrictions on possible six-dimensional $(2, 0)$ theories. We will in fact find that they obey an $ADE$-classification, i.e. they are in one-to-one correspondence with the discrete subgroups of $SU (2)$, or the simply laced Lie groups $SU (r + 1)$, $SO (2 r)$, $E_6$, $E_7$, and $E_8$. This result was indeed predicted by the original definition of the $(2, 0)$ theories in terms of ten-dimensional type IIB string theory on the product of $M$ and a four-manifold with a simple singularity \cite{Witten95}. However, it is gratifying to recover it by a purely six-dimensional argument. The $(2, 0)$ theories may also be regarded as the six-dimensional origin of certain $N = 4$ super Yang-Mills theories in four space-time dimensions, and from this point of view, it is of course more surprising that only models with a simply laced gauge group can appear. 

Because of its electromagnetic charge, a `bare' string is always surrounded by an electromagnetic field configuration, and as mentioned above, the classical anomaly arises from the strong interaction of the string with this self-field. In view of this, it is natural to try to define a `dressed' string, which only interacts weakly with other excitations of the theory. A concrete situation to consider is a configuration with a static, straight string. The tension of the string is given by the vacuum expectation value of the scalar field $\phi$. At energies low compared to the square root of this tension, the theory consists of a Minkowski space field theory (the free tensor multiplet theory) and a decoupled theory describing fluctuations of the dressed string. In section four, we will show that the latter world-sheet theory is a supersymmetric version of the level one Wess-Zumino-Witten model based on the group $({\mathbb R} \times SU (2)) / {\mathbb Z}_2$. However, we would like to caution the reader that the results of this section are less well established than those in the preceeding sections.

\section{Classical and quantum anomalies}
\subsection{The descent formalism}
On the two-dimensional world-sheet $\Sigma$, an anomaly under gauge transformations that can be continuously connected to the identity (so called `perturbative anomalies' as opposed to `global anomalies') is described by a four-dimensional integer characteristic class $I$. The two extra dimensions arise from the necessity of considering a two-parameter family of gauge-field configurations \cite{Atiyah-Singer}\cite{Zumino}. The gauge invariant four-form $I$ is closed. Locally, it can thus be written as $I = d \omega$ for some three-form $\omega$, which however will not be gauge invariant. Its variation under an infinitesimal gauge transformation is of the form $\delta \omega = d {\cal A}$, where the two-form ${\cal A}$ is linear in the parameters of the transformation. The anomalous variation of the effective action is then given by $2 \pi \int_\Sigma {}^* ({\cal A})$.

\subsection{The electric coupling}
To exhibit the anomaly of the electric coupling in (\ref{action}), we can rewrite it in various ways, none of which is completely satisfactory, though, but should rather be seen as heuristic expressions:
\beq
e \int_\Sigma {}^* \, (b) = e \int_M b \wedge \delta_\Sigma = e \int_M h^{\rm tot} \wedge \delta_D = e \int_D {}^* \, (h^{\rm tot}) . \label{electric}
\eeq
Starting from the left, the first definition is given by integrating the pullback of $b$ over the world-sheet $\Sigma$. In the second version, we have used the Poincar\'e dual $\delta_\Sigma$ of $\Sigma$ to rewrite it as an integral over space-time $M$. \footnote{The four-form $\delta_\Sigma$ is defined by the property that $\int_M \delta_\Sigma \wedge s = \int_\Sigma {}^* (s)$ for an arbitary test-function two-form $s$. An explicit expression is $\delta_\Sigma = dx^\mu \wedge dx^\nu \wedge dx^\rho \wedge dx^\sigma \int_\Sigma d X^\tau \wedge d X^\kappa \delta^{(6)} (x - X) \epsilon_{\mu \nu \rho \sigma \tau \kappa}$.} The problem with these expressions is that they are not manifestly gauge-invariant, since they depend on the non gauge-invariant quantity $b$. To remedy this, we can try to work solely with the gauge-invariant field strength $h^{\rm tot}$ as in the third expression. To incorporate the modified Bianchi identity (\ref{Bianchi}) we define $h^{\rm tot} = d b + 2 \pi q \, \delta_D$, where $\delta_D$ is a three-form such that $d \delta_D = \delta_\Sigma$. Note that $\delta_D \wedge \delta_D = 0$. But no matter how we choose $\delta_D$, such a choice partly breaks the symmetry under diffeomorphisms of $M$ that leave $\Sigma$ invariant. A particular choice, which is used in going to the last expression, is to take $\delta_D$ as the Poincar\'e dual of an open three-manifold $D$, the boundary of which is given by $\Sigma$. (This Poincar\'e dual is defined in complete analogy with $\delta_\Sigma$.) In this case, the unbroken subgroup consists of diffeomorphisms that leave $D$ invariant. These difficulties are what we have in mind when we say that the electric coupling is anomalous. 

Under an infinitesimal gauge transformation, the variation of $\delta_D$ is of the form $\delta (\delta_D) = \frac{1}{q e} d {\cal A}$, for some two-form ${\cal A}$ which is linear in the parameters of the transformation. The variation of the electric coupling, as given by the third expression in  (\ref{electric}), is then 
\beq
\frac{e}{q e} \int_M h^{\rm tot} \wedge d {\cal A} = \frac{1}{q} \int_M d h^{\rm tot} \wedge {\cal A} = 2 \pi \int_M \delta_\Sigma \wedge {\cal A} = 2 \pi \int_\Sigma {}^* ({\cal A}) ,
\eeq
where we have used the  modified Bianchi identity (\ref{Bianchi}). So this would seem to fit into the descent formalism, if we take the characteristic class $I$ to equal $q e \, {}^* (\delta_\Sigma)$. Indeed, this means that we can take $\omega = q e \, {}^* (\delta_D)$ so that $\delta (\omega) = d {\cal A}$ as required. As can be most clearly seen from the second expression in (\ref{electric}), this is in fact a particular case of anomaly inflow from the six-dimensional bulk to the two-dimensional world-sheet, with the only unfamiliar feature being that the anomaly four-form is given by the Poincar\'e dual $\delta_\Sigma$. \footnote{As discussed above, one should really consider a two-parameter family of world-sheets embedded in a two-parameter family of space-times. We will not make this explicit, though, but one should bear in mind that ${}^*$ in these formulas denote the pullback to this two-parameter family of world-sheets rather than to $\Sigma$.}

Since $\delta_\Sigma$ has delta-function support on $\Sigma$, taking its pullback ${}^* (\delta_\Sigma)$ might appear as a rather singular operation, but it actually has a well-defined meaning, as we will now explain: In a tubular neighborhood of $\Sigma$, we can approximate space-time $M$ with the total space of the normal bundle $N$. The Poincar\'e dual $\delta_\Sigma$ then defines a class $\Phi$ in the cohomology with compact vertical support $H^4_v (N)$ on this space. In fact, $\Phi$ is the Thom class, i.e. the image of $1$ under the Thom isomorphism $H^0 (\Sigma) \simeq H^4_v (N)$. We are thus interested in the pullback of $\Phi$ by the zero section of $N$. But a general theorem states that this equals the Euler class $\chi (N)$ of the normal bundle. (These matters are explained in more detail in textbooks on algebraic topology, see e.g. Section I.6 of \cite{Bott-Tu} or Chapter 21 of \cite{Madsen-Tornehave}.) 

So we find that the characteristic class $I^{\rm class}$ associated with the classical anomaly of the electric coupling is given by
\beq
I^{\rm class} = q e \, \chi (N) .
\eeq 

\subsection{The chiral fermions}
As described in the introduction, the chiral fermions $\Theta^+$ and $\Theta^-$ are sections of $\Sigma^+ \otimes N^+ \otimes E^+$ and $\Sigma^- \otimes N^- \otimes E^-$ respectively. Here, $\Sigma^\pm$, $N^\pm$, and $E^\pm$ are the positive and negative chirality spinor bundles associated with the world-sheet tangent bundle $T \Sigma$, the normal bundle $N$, and the $R$-symmetry bundle $E$ respectively. The contributions to the anomalies now follow from standard formulas: For $\Theta^+$ and $\Theta^-$ we get $ch_2 (N^+)$ and $- ch_2 (N^-)$ respectively, where $ch_2 (N^+)$ and $ch_2 (N^-)$ denote the second Chern character classes. (A factor $\frac{1}{2}$ due to the reality conditions on $\Theta^+$ and $\Theta^-$ cancels against a factor $2$ corresponding to the rank of the bundles $E^+$ and $E^-$.)

The classes $ch_2 (N^+)$ and $ch_2 (N^-)$ are related to the Euler class $\chi (N)$ and the first Pontryagin class $p_1 (N)$:
\bea
\chi (N) & = & ch_2 (N^-) - ch_2 (N^+) \cr
p_1 (N)  & = & ch_2 (N^-) + ch_2 (N^+) . \label{classes-relations}
\eea
These relationships reflect the fact that
\beq
SO (4) \simeq Spin (4) / {\mathbb Z}_2 \simeq (SU (2) \times SU (2)) / {\mathbb Z}_2 , \label{SO4}
\eeq
where $SO (4)$ is the structure group of $N$, and the two $SU (2)$ factors are the structure groups of $N^+$ and $N^-$ respectively. Attempting to invert these relationships to express the integer classes $ch_2 (N_+)$ and $ch_2 (N^-)$ in terms of $\chi (N)$ and $p_1 (N)$, we find that the mod $2$ reduction of $\chi (N) + p_1 (N)$ is the obstruction to lifting the $SO (4)$ bundle $N$ to a $Spin (4)$ bundle.

So we find that the characteristic class $I^{\rm quant}$ associated with the one-loop quantum anomaly of the chiral fermions is given by
\beq
I^{\rm quant} = - \chi (N) .
\eeq

\section{The ADE-classification}
\subsection{The coupling constant}
Before discussing the possibilities for the anomalies found in the previous section to cancel, we will determine the correct value of the coupling constant $\lambda$, that appears in the action (\ref{action}). In the absence of charged strings, the field strength $h^{\rm tot}$ is closed, and we have normalized it so that $\frac{1}{2 \pi} h^{\rm tot}$ is a representative of an integer class. The replacement $\lambda \rightarrow 1 / \lambda$ then defines an equivalent theory. This is completely analogous to the $S$-duality of four-dimensional Maxwell theory, or the $T$-duality of a compact boson in two dimensions. Only the self-dual part of $h^{\rm tot}$ is part of the tensor multiplet, though, and for an irrational value of $\lambda^2$, we do not know how to define the corresponding quantum theory. But for a rational value of $\lambda^2$, the Hilbert space of $h^{\rm tot}$ is a finite sum (with the number of terms depending on the value of $\lambda^2$), where each term is a tensor product of two Hilbert spaces, pertaining to the self-dual and anti self-dual parts respectively. For a single free tensor multiplet, i.e. the world-volume theory of a single five-brane in $M$-theory, it appears that the correct value to use is $\lambda^2 = 2$, analogous to the `free fermion radius' for a boson in two dimensions. Indeed, the additional topological data provided by the embedding into $M$-theory is then precisely what is needed to pick out the correct term in this sum \cite{Witten96}. The value $\lambda^2 = 2$ can also be determined by the requirement that observables associated with different closed spatial surfaces commute with each other \cite{Henningson}. As we mentioned in the introduction, for this value of $\lambda$, the decoupling of the anti self-dual part of the field strength requires that the electric and magnetic charges $e$ and $q$ of a string are related as $e = \frac{1}{2} q$.  

\subsection{Anomaly cancellation}
Sofar, we have been discussing a theory with a single tensor multiplet and a single type of string. We now generalize this by introducing a tensor multiplet that takes its values in a real vector space $W$ endowed with a positive definite inner product denoted as $w \cdot w^\prime$ for $w, w^\prime \in W$. (The positivity requirement of the inner product is necessary for the positivity of the Hamiltonian.) We also introduce a discrete subset $Q \subset W$ of allowed magnetic string charges $q$. 

We will now determine all possible such sets $Q$ of allowed magnetic charges. A first restriction is obtained by considering a single string with magnetic charge $q \in Q$, and thus electric charge $e = \frac{1}{2} q \in W$. From our results in the previous section, it follows that the total anomaly of such a string, taking both classical and quantum contributions into account, is given by the characteristic class $I = I^{\rm class} + I^{\rm quant} = \frac{1}{2} (q \cdot q - 2) \chi (N)$. All other terms in the action (\ref{action}) are non-anomalous. Cancellation of the anomaly thus requires that
\beq
q \cdot q = 2 , \label{length}
\eeq
for all $q \in Q$. \footnote{This condition ensures that perturbative anomalies cancel. One should then continue and also investigate possible global anomalies, but we will not pursue this issue in the present paper.} 

\subsection{Dirac quantization}
Another restriction on the spectrum of allowed charges follows from considering two different strings with electric and magnetic charges $(e, q)$ and $(e^\prime, q^\prime)$ respectively. (We temporarily relax the relationship between electric and magnetic charges.) To begin with, we suppose that the electric charge of the first string and the magnetic charge of the second string vanish, i.e. $e = 0$ and $q^\prime = 0$. Because of the magnetic charge $q$ of the first string, there is a non-trivial magnetic field $h^{\rm tot}$ in space-time. On the complement $M^* = M - \Sigma$ of the world-sheet $\Sigma$ of the first string, $h^{\rm tot}$ is closed and defines a cohomology class $[h^{\rm tot}] = q \Omega$, where $\Omega$ is an element of $H^3 (M^*, {\mathbb Z})$. Because of the electric charge $e^\prime$ of the second string, the quantum `wave function' $\Psi$ of this system is not a complex function but rather a section of a complex line-bundle ${\cal L}$ over the configuration space ${\cal M}$. (This is the space of all configurations in which the two strings do not intersect each other.) This line-bundle is completely characterized by its Chern class $c_1 ({\cal L})$, which we can specify by evaluating it on all possible two-cycles $s \in H_2 ({\cal M}, {\mathbb Z})$. Such a two-cycle $s$ defines a three-cycle $S \in H_3 (M^*, {\mathbb Z})$, and we have that $\int_s c_1 ({\cal L}) = e^\prime \cdot q \int_S \Omega$. The generalization to arbitrary charges $(e, q)$ and $(e^\prime, q^\prime)$ is that $\int_s c_1 ({\cal L}) = (e^\prime \cdot q + e \cdot q^\prime) \int_S \Omega$. 

One should note the relative plus sign between the terms, as opposed to the minus sign familiar from the theory of dyonic particles in four dimensions. As explained in \cite{Deser-Gomberoff-Henneaux-Teitelboim}, this difference can be understood by a careful consideration of the topology of the configuration space ${cal M}$. Heuristically, it is related to the fact that the wedge product of two-forms in four dimensions is symmetric, whereas the wedge product of three-forms in six dimensions is anti-symmetric. This sign has profound consequences, though: In four dimensions, we usually do not allow the world-lines of mutually non-local dyons to intersect. The world-line of a single dyon of course 'intersects' itself, but this does not lead to any problems, since the minus sign ensures that dyons of the same kind are mutually local. In six dimensions, however, the plus sign means that a dyonic string is not mutually local with itself, and this can be seen as the origin of the classical bosonic anomaly discussed in the previous section.

From the integrality of the class $c_1 ({\cal L})$, it now follows that the number $e^\prime \cdot q + e \cdot q^\prime$ must be an integer. Reinstating the relationship between electric and magnetic charges, e.g. $e = \frac{1}{2} q$ and $e^\prime = \frac{1}{2} q^\prime$, we thus find that
\beq
q \cdot q^\prime \in {\mathbb Z} \label{angle}
\eeq
for all $q, q^\prime \in Q$.

\subsection{The $ADE$-classification}
The conditions (\ref{length}) and (\ref{angle}) on the elements of $Q$ are precisely those that define the roots of a simply laced Lie algebra, i.e. the algebras $A_r \simeq su (r + 1)$ for $r = 1, 2, \ldots$, $D_r \simeq so (2 r)$ for $r = 4, 5, \ldots$, and $E_r$ for $r = 6, 7, 8$. We have thus recovered the $ADE$-classification of consistent $(2, 0)$ theories by a purely six-dimensional argument. 

This means that we should think of the tensor multiplet as taking its values in the weight space $W$ of the corresponding simply laced Lie algebra. The set $Q$ of allowed magnetic charges can then be defined as 
\beq
Q = \left\{ q \in \Gamma^r \Bigl| q \cdot q = 2 \right\} ,
\eeq
where $\Gamma^r$ is the root lattice of this algebra. The three-form field strength $h^{\rm tot}$ is subject to the restriction that its periods $\frac{1}{2 \pi} \int h^{\rm tot}$, where the integral is taken over a three-cycle in $M$, should be elements of the weight lattice $\Gamma^w \subset W$, i.e. the dual of the root lattice $\Gamma^r \subset W$. 

Locally, $h^{\rm tot} = d b$ for some two-form $b$, subject to gauge transformations of the form $b \rightarrow b + \Delta b$. The parameter $\Delta b$ is a closed two-form whose periods $\frac{1}{2 \pi} \int \Delta b$, where the integral is taken over a two-cycle in $M$, are elements of $\Gamma^w$. The theory should be invariant under such transformations. However, the factor $\frac{1}{2}$ in the relationship $e = \frac{1}{2} q$ implies that the exponentiated electric coupling $\exp \left(i \int_\Sigma e \, {}^* (b) \right)$ will in general only be invariant up to a sign. The theory thus appears to suffer from a global anomaly under such gauge transformations. (There is no perturbative anomaly, since the coupling is invariant under transformations with an exact parameter $\Delta b$.) Hopefully, this anomaly is cancelled by a similar sign ambiguity in the definition of the path integral measure for the fermions $\Theta^+$ and $\Theta^-$, but we will not attempt to show this in the present paper. 

\subsection{The $A_r$-model}
It is instructive to consider the $A_r$ model in somewhat more detail. It can be realized as the world-volume theory of $r + 1$ parallell five-branes in $M$-theory, each of which supports a tensor multiplet. Membranes stretching from one five-brane to another appear as $r (r + 1)$ different types of strings in six dimensions. Their magnetic charges with the respect to the tensor multiplets are given by vectors $q$ with $r + 1$ entries of the form 
\beq
q = (0, \ldots, 0, +1, 0, \ldots, 0, -1, 0, \ldots, 0) .
\eeq
Indeed the set $Q$ of such charges fulfills the conditions (\ref{length}) and (\ref{angle}).

One linear combination of the $r + 1$ tensor multiplets, namely the sum, decouples from all the strings. This is analogous to the world-volume theory of $r + 1$ D3-branes in type IIB string theory, where the gauge group is $SU (r + 1)$ rather than $U (r + 1)$, because a central $U (1)$ factor locally decouples from the rest of the theory. We should thus focus our attention on the $r$-dimensional linear space $W$ orthogonal to the sum of the tensor multiplets. This can naturally be identified with the weight space of the $A_r \simeq su (r + 1)$ Lie algebra, and $Q$ is then indeed the set of roots of this algebra. 

For a single tensor multiplet, the periods of $h^{\rm tot}$ take their values in ${\mathbb Z}$, so for $r + 1$ tensor multiplets, we instead get ${\mathbb Z}^{r + 1}$. But an element $k = (k_1, \ldots, k_{r + 1})$ of the latter group may be decomposed as $k = s (1, \ldots, 1) + w$, where $s$ is some multiple of $\frac{1}{r + 1}$, and $w$ belongs to the $A_r$ weight lattice $\Gamma^w$ defined as
\beq
\Gamma^w = \left\{ (w_1, \ldots, w_{r + 1}) \in \frac{1}{r \! + \! 1} {\mathbb Z} \times \ldots \times \frac{1}{r \! + \! 1} {\mathbb Z} \Bigl|   w_i - w_j \in {\mathbb Z}, w_1 + \ldots + w_{r + 1} = 0 \right\} .
\eeq
The root lattice $\Gamma^r$, i.e the dual of $\Gamma^w$, is then given by
\beq
\Gamma^r = \left\{ (q_1, \ldots, q_{r + 1}) \in {\mathbb Z} \times \ldots \times {\mathbb Z} \Bigl| q_1 + \ldots + q_{r + 1} = 0 \right\} .
\eeq

\section{The decoupled world-sheet theory}
Sofar we have been describing the theory in terms of `bare' string, and certain space-time fields $\phi$, $\psi$, and $h^{\rm tot}$. But for some purposes, these are not the most convenient variables to use. The reason is, that in the presence of a string, a configuration with vanishing space-time fields is not possible. Indeed, the modified Bianchi identity (\ref{Bianchi}) is an example of this phenomenon. 

We would therefore like to change variables, and describe the theory in terms of fluctuations around a configuration with a `dressed' string, which includes this self-field. A concrete situation where this would be useful, is a configuration containing a straight, static string. The string is characterized by its tension, which is given by the vacuum expectation value of the scalar field $\sqrt{\phi \phi}$. At energies low compared to the scale set by (the square root of) the tension, we expect that the theory factorizes into a space-time sector and a world-sheet sector, that are weakly coupled to each other. In the infra-red limit, we expect them to decouple completely. (It is more convenient to describe the decoupling limit by considering fluctuations of some fixed wavelength while taking the string tension to infinity.) The space-time sector is then of course the theory of a free tensor multiplet, and the world-sheet theory must be some two-dimensional conformal field theory. 

\subsection{The spherical variables}
To formulate this decoupled world-sheet theory, we start by considering the normal bundle $N$ of the world-sheet $\Sigma$ embedded in space-time $M$. As described in the introduction, the world-sheet field $X^\perp$ is a section of this bundle. We may rewrite $N$ as
\beq
N \simeq (R \times S) / {\mathbb Z}_2 ,
\eeq
where $R$ is a real line bundle with fiber ${\mathbb R}$, and $S$ is a three-dimensional sphere bundle with fiber $S^3 \simeq SU (2)$. The non-trivial element of ${\mathbb Z}_2$ acts by multiplication with $-1$ on ${\mathbb R}$ and as the antipodal map on $S^3$ (i.e. by multiplication with the non-trivial element of the center of $SU (2)$). This of course corresponds to writing $X^\perp$ in terms of a radius $r \in {\mathbb R}$, which is a section of $R$, and three angular variables $g \in SU (2)$, that constitute a section of $S$. The restriction to positive $r$, which is customary in spherical coordinates, is replaced by the ${\mathbb Z}_2$ equivalence relation. 

In this formulation, the $SO (4) \simeq (SU (2) \times SU (2)) / {\mathbb Z}_2$ structure group of $N$ acts as
\bea
r & \mapsto & r \cr
g & \mapsto & u g v^{-1} , \label{rgtransf}
\eea
for $u, v \in SU (2)$. Its action on the fermionic fields $\Theta^+$ and $\Theta^-$ is
\bea
\Theta^+ & \mapsto u \Theta^+ \cr
\Theta^- & \mapsto v \Theta^- . \label{Thetatransf}
\eea
The fermionic fields are also doublets under the unbroken $SO (4) \subset SO (5)_R$ symmetry and obey a reality condition, but we will suppress these structures from our notation. The covariant exterior derivative $D$ acting on the various fields is thus
\bea
D r & = & d r \cr
D g & = & d g + A g - g \tilde{A} \cr
D \Theta^+ & = & (d + A) \Theta^+ \cr
D \Theta^- & = & (d + \tilde{A}) \Theta^- ,
\eea
where the $SO (4)$ connection on $N$ is expressed as a pair of $SU (2)$ connections $A$ and $\tilde{A}$ transforming as
\bea
A & \mapsto & u A u^{-1} - d u u^{-1} \cr
\tilde{A} & \mapsto & v \tilde{A} v^{-1} - d v v^{-1} . \label{AtildeAtransf}
\eea
As usual, the corresponding covariant field strengths are defined as $F = d A + A \wedge A$ and $\tilde{F} =  d \tilde{A} + \tilde{A} \wedge \tilde{A}$.

\subsection{The gauged Wess-Zumino term}
We now wish to construct the world-sheet theory describing fluctuations around a configuration with a straight, static string of infinite tension. The action is in fact largely determined by the anomaly structure described in the previous sections, and must take the form
\beq
S = \int_\Sigma d^2 \sigma \, {\cal L} + S_{WZ} . \label{wsaction}
\eeq
Here ${\cal L}$ is some gauge invariant local Lagrangian density, and $S_{WZ}$ is a non-local gauged Wess-Zumino term. To define the latter, we must extend the domain of definition of the field $g$ from $\Sigma$ to an open three-manifold $D$, the boundary of which equals $\Sigma$. We then have
\bea
S_{WZ} & = & \frac{1}{4 \pi} \int_\Sigma {\rm Tr} \left( g^{-1} d g \tilde{A} + d g g^{-1} A - g \tilde{A} g^{-1} A \right) \cr
& & +\frac{1}{12 \pi} \int_D {\rm Tr} \left( g^{-1} d g \wedge g^{-1} dg \wedge g^{-1} d g \right) ,  
\eea
where ${\rm Tr}$ denotes the trace in the fundamental representation of $SU (2)$. Since the integrand of the second term equals $2 \pi$ times the generator of $H^3 (S^3, {\mathbb Z})$, $S_{WZ}$ is a level one gauged Wess-Zumino term. It is a well-defined functional mod $2 \pi$ of $A$, $\tilde{A}$, and the restriction of $g$ to $\Sigma$. The reason for including it in the action (\ref{wsaction}) is that it has the correct anomalous transformation properties under (\ref{rgtransf}) and (\ref{AtildeAtransf}). Indeed, a short calculation shows that
\beq
S_{WZ} \mapsto S_{WZ} + \frac{1}{4 \pi} \int_\Sigma {\rm Tr} \left(u^{-1} du A - v^{-1} dv \tilde{A} \right) \;\;\; {\rm mod} \;\; 2 \pi .
\eeq
For $u$ and $v$ infinitesimally close to the unit element of $SU (2)$, the anomalous variation of $S_{WZ}$ follows by applying the descent procedure to the characteristic class
\beq
I^{\rm class} = ch_2 (F) - ch_2 (\tilde{F}) = \frac{1}{8 \pi^2} {\rm Tr} (F \wedge F) - \frac{1}{8 \pi^2} {\rm Tr} (\tilde{F} \wedge \tilde{F}) .
\eeq
As discussed in section two, this coincides with the classical anomaly of the electric coupling for a string with magnetic charge $q$ such that $q \cdot q = 2$ and electric charge $e = \frac{1}{2} q$.

\subsection{The local terms}
It remains to determine the local Lagrangian density ${\cal L}$ in (\ref{wsaction}). But this follows from various symmetry requirements, notably the conformal invariance of the model. The universality class of the model is in fact governed by the gauged Wess-Zumino term $S_{WZ}$ described in the previous subsection. We find that ${\cal L}$ is sum of separate kinetic terms for the fields $r$, $g$, $\Theta^+$, and $\Theta^-$.

Up to a field redefinition, the radial field $r$ is a free non-compact boson with Lagrangian density
\beq
{\cal L}_r = - \frac{1}{4 \pi} D_+ r D_- r .
\eeq
For the angular field $g$, we have a gauged non-linear sigma-model Lagrangian density
\beq
{\cal L}_g = \frac{1}{4 \pi} {\rm Tr} \left( g^{-1} D_+ g g^{-1} D_- g \right) .
\eeq
Finally, the fermionic fields $\Theta^+$ are $\Theta^-$ are governed by the gauged Dirac Lagrangian density
\beq
{\cal L}_\Theta = \frac{i}{4 \pi} \left( (\Theta^+)^\dagger D_+ \Theta^+ + (\Theta^-)^\dagger D_- \Theta^- \right) .
\eeq
In these formulas, $D_+$ and $D_-$ denote the covariant derivatives with respect to the world-sheet light-cone coordinates $\sigma^+$ and $\sigma^-$. The last term gives a one-loop contribution $I^{\rm quant}$ to the anomaly of the quantum effective action, as described in section two. 

The normalizations of ${\cal L}_r$ and ${\cal L}_\Theta$ are conventional, but the coefficient in front of ${\cal L}_g$ is significant. One could imagine that its `bare' value is very large, corresponding to the large tension of the string. This means that the sigma-model is weakly coupled. Let us now consider a configuration with $A = \tilde{A} = 0$ as is appropriate for a static, straight string. As described in \cite{Witten84}, the coupling constant of ${\cal L}_g$ will then flow to a non-trivial infra-red fixed point determined by the coefficient of the Wess-Zumino term $S_{WZ}$. The critical value can be most easily described by giving the variation of the total action (\ref{wsaction}) under a general variation $\delta g$ of the field $g$:
\beq
\delta S = \frac{1}{4 \pi} \int_\Sigma d^2 \sigma {\rm Tr} \left( g^{-1} \delta g \partial_+ (g^{-1} \partial_- g) \right) = \frac{1}{4 \pi} \int_\Sigma d^2 \sigma {\rm Tr} \left( \delta g g^{-1} \partial_- (\partial_+ g g^{-1}) \right) ,
\eeq
i.e. the chiral currents $J_- = g^{-1} \partial_- g$ and $J_+ = \partial_+ g g^{-1}$ are separately conserved at the critical point.

\subsection{World-sheet supersymmetry}
A static straight string configuration breaks half of the supersymmetries of the six-dimensional $(2, 0)$ supersymmetry algebra. The broken symmetries have infinitesimal parameters $\lambda^+$ and $\lambda^-$ with the same quantum numbers as the Goldstino fields $\Theta^+$ and $\Theta^-$. They act non-linearly on the fields:
\bea
\delta r & = & 0 \cr
\delta g & = & 0 \cr
\delta \Theta^+ & = & \lambda^+ \cr
\delta \Theta^- & = & \lambda^- .
\eea
The unbroken symmetries have infinitesimal parameters $\eta^+$ and $\eta^-$ with quantum numbers that differ from those of $\Theta^+$ and $\Theta^-$ in that the world-sheet chiralities are reversed. They act linearly on the fields:
\bea
- i g^{-1} \delta g + {\mathbb I} \delta r & = & \Theta^+ (\eta^+)^\dagger + \eta^+ (\Theta^+)^\dagger \cr
\delta \Theta^+ & = & (g^{-1} \partial_- g + i {\mathbb I} \partial_- r) \eta^+ \cr
\delta \Theta^- & = & 0 
\eea
for the $\eta^+$ transformations, and
\bea
- i \delta g g^{-1} + {\mathbb I} \delta r & = & \Theta^- (\eta^-)^\dagger + \eta^- (\Theta^-)^\dagger \cr
\delta \Theta^+ & = & 0 \cr
\delta \Theta^- & = & (\partial_+ g g^{-1} + i {\mathbb I} \partial_+ r) \eta^-  
\eea
for the $\eta^-$ transformations. Here ${\mathbb I}$ is the $2 \times 2$ unit matrix. A straightforward computation shows that the total action (\ref{wsaction}) is invariant under these transformations when $A = \tilde{A} = 0$.

\vspace*{5mm}
This work was inspired by a visit to the Institute for Advanced Study in Princeton. I would like to thank the Institute for its hospitality, and Edward Witten for drawing my attention to the issue of anomalies. I am supported by a Research Fellowship from the Royal Swedish Academy of Sciences (KVA).

\end{document}